\documentclass[sigconf]{acmart}
\usepackage{longtable}
\usepackage{booktabs}
\usepackage{algorithmic}
\usepackage{graphicx}
\usepackage{subfigure}
\usepackage{textcomp}
\usepackage{booktabs}
\usepackage{amsmath,amscd,amsbsy,amssymb,amsfonts,latexsym,url,bm,amsthm}
\def\BibTeX{{\rm B\kern-.05em{\sc i\kern-.025em b}\kern-.08em
		T\kern-.1667em\lower.7ex\hbox{E}\kern-.125emX}}
\usepackage[vlined,boxed,commentsnumbered,linesnumbered,ruled]{algorithm2e}

\usepackage{threeparttable}

\copyrightyear{Proceedings of the 2019 World Wide Web Conference, published under Creative Commons CC-BY 4.0 License.}
\acmYear{2019} 
\acmConference[WWW '19]{Proceedings of the 2019 World Wide Web Conference}{May 13--17, 2019}{San Francisco, CA, USA}
\acmPrice{}
\acmDOI{10.1145/3308558.3313442}
\acmISBN{978-1-4503-6674-8/19/05}

\begin{document}
\title{Dual Graph Attention Networks for Deep Latent Representation of Multifaceted Social Effects in Recommender Systems}

\author{Qitian Wu$^1$, Hengrui Zhang$^1$, Xiaofeng Gao$^1$\footnote{Xiaofeng Gao is the corresponding author. This work was supported by the National Key R\&D Program of China [2018YFB1004703]; the National Natural Science Foundation of China [61872238, 61672353]; the Shanghai Science and Technology Fund [17510740200]; the Huawei Innovation Research Program [HO2018085286]; and the State Key Laboratory of Air Traffic Management System and Technology [SKLATM20180X].}, Peng He$^2$, Paul Weng$^3$, Han Gao$^2$, Guihai Chen$^1$}
\orcid{1234-5678-9012}
\affiliation{%
	\institution{$^1$Shanghai Key Laboratory of Scalable Computing and Systems,\\ Department of Computer Science and Engineering, Shanghai Jiao Tong University\\
	$^2$WeChat, Tencent Inc.\\
	$^3$UM-SJTU Joint Institute, Shanghai Jiao Tong University\\}
}
\email{{echo740, sqstardust}@sjtu.edu.cn, gao-xf@cs.sjtu.edu.cn, paulhe@tencent.com,} \email{paul.weng@sjtu.edu.cn, alangao@tencent.com,
	gchen@cs.sjtu.edu.cn}

\thanks{*Xiaofeng Gao is the corresponding author. This work was supported by the National Key R\&D Program of China [2018YFB1004703]; the National Natural Science Foundation of China [61872238, 61672353]; the Shanghai Science and Technology Fund [17510740200]; the Huawei Innovation Research Program [HO2018085286]; and the State Key Laboratory of Air Traffic Management System and Technology [SKLATM20180X]; the Tencent Social Ads Rhino-Bird Focused Research Program}

\begin{comment}
\author{\IEEEauthorblockN{Qitian Wu$^1$, Hengrui zhang$^1$, Xiaofeng Gao$^1$\footnote{Xiaofeng Gao is the corresponding author.}, Peng He$^2$, Paul Weng$^1$, Han Gao$^2$, Guihai Chen$^1$}
	$^1$Shanghai Jiao Tong University, China\\
	$^2$WeChat, Tencent Inc.\\
	\{echo740, sqstardust\}@sjtu.edu.cn, gao-xf@cs.sjtu.edu.cn, paulhe@tencent.com, paul.weng@sjtu.edu.cn, alangao@tencent.com,
	gchen@cs.sjtu.edu.cn
}
\thanks{*Xiaofeng Gao is the corresponding author.}
\end{comment}

\begin{abstract}
	Social recommendation leverages social information to solve data sparsity and cold-start problems in traditional collaborative filtering methods. 
	However, most existing models assume that social effects from friend users are static and under the forms of constant weights or fixed constraints. 
	To relax this strong assumption, in this paper, we propose \emph{dual graph attention networks} to collaboratively learn representations for two-fold social effects, where one is modeled by a user-specific attention weight and the other is modeled by a dynamic and context-aware attention weight.
	We also extend the social effects in user domain to item domain, so that information from related items can be leveraged to further alleviate the data sparsity problem. 
	Furthermore, considering that different social effects in two domains could interact with each other and jointly influence users' preferences for items, we propose a new policy-based fusion strategy based on \emph{contextual multi-armed bandit} to weigh interactions of various social effects. 
	Experiments on one benchmark dataset and a commercial dataset verify the efficacy of the key components in our model. The results show that our model achieves great improvement for recommendation accuracy compared with other state-of-the-art social recommendation methods.
\end{abstract}

%
% The code below should be generated by the tool at
% http://dl.acm.org/ccs.cfm
% Please copy and paste the code instead of the example below.
%

\maketitle

\section{Introduction}

Recommender systems, which aim at filtering and suggesting items of potential interests to targeted users, have extensive applications and can bring up a huge amount of commercial benefits, especially when the information overload problem has been pervasive with the rapid development of the World Wide Web. 
Conventional collaborative filtering (CF) methods for recommendation suffer from data sparsity and cold start problems. 
To this end, social recommendation, which attempts to leverage the social networks of users to incorporate more useful information, has been proposed to alleviate these two problems \cite{Trust1,Trust2,SocReg,SOREC}. 

\textbf{Prior Works and Limitations.} Nowadays, there are many composite platforms which integrate item recommendation and social network services (SNS). 
Users on these platforms can click on or rate items as well as interact with their friends. 
A user's preference for one item could be impacted by her friends, which motivates us to probe into such social effect to improve recommendation quality. 
Previous studies for social recommendation attempt to model social effects in various ways, such as by trust propagation  \cite{Trust1,Trust2,Trust3}, regularization loss \cite{SocReg,SocialMF}, matrix factorization \cite{MF1,MF2,TrustMF}, network embedding \cite{SREPS,NetRep1,DBLP:conf/cikm/ZhangLNLX18}, and deep neural network \cite{Deep1,NSCR}. 

\begin{figure}[h]
	\centering
	\includegraphics[width=0.47\textwidth,angle=0]{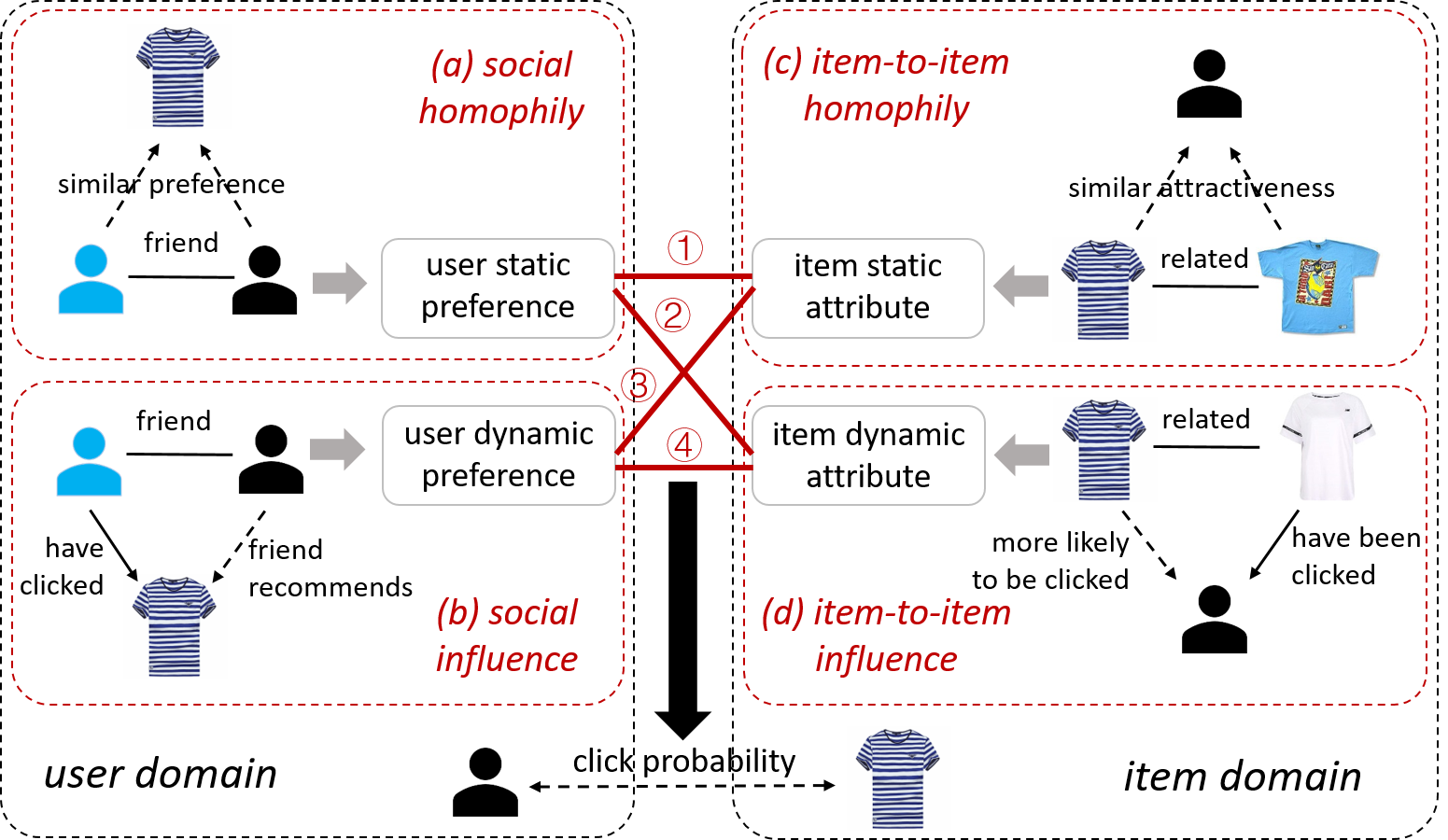}
	\vspace{-5pt}
	\caption{Illustration of the two-fold social effects, i.e., homophily effect and influence effect, in user social networks as well as among related items. The four social effects jointly affect a user's decision on one item.}
	\label{fig-motivation}
	\vspace{-5pt}
\end{figure}

However, these studies share several common limitations. \textbf{First}, most studies assume that linked users all share similar preferences. This assumption cannot suit well contemporary SNS since there could be various types of online friends, such as close friends, casual friends, and event friends \cite{Friends}. 
Therefore, {directly equating} friendship on an SNS with preference similarity may not be sound. 
\textbf{Second}, most works model friends' influences statically under the forms of constant weights or fixed constraints. 
This assumption ignores the dynamic pattern of social effects. 
In fact, users could be influenced by a specific group of friends when it comes to specific items, which makes the social effect dynamic and dependent of specific contexts. 
\textbf{Third}, previous methods lack interpretability for social effects, i.e., they cannot explicitly indicate how a user's preference for one item is influenced by their friends.

\textbf{Motivations and Rationales.} 
%WeChat is a burgeoning SNS in China and has over 1 billion daily active users. The recently released application Top Story on WeChat aims at recommending articles published by WeChat Official Account to users. WeChat users can click and read articles in Top Story, communicate with friends via texts or speeches, and browse friends' posted contents (including texts, articles, musics or videos, etc) or post contents in WeChat Moment (a community formed by users in ego-network). Fig. \ref{fig-kanyikan} shows the UI surfaces for three situations. Frequent interactions with friends via messages and WeChat Moments could influence users' preference and their clicking behaviors.
In the real world, people's decisions could be influenced by various factors, and their behaviors are often the results of multifaceted causes. 
To this end, we probe into four different social effects in recommender systems, including two-fold effects in user domain and the other symmetric two-fold ones in item domain, as illustrated in Fig.~\ref{fig-motivation}. 
In user domain,
friends could influence each other in two ways. 
For one thing, users tend to possess similar preferences to their friends, which is called as \emph{social homophily} \cite{Homophily} (Fig.~\ref{fig-motivation}.(a)). 
Social homophily often leads to intrinsic effect for user preference which stays unchanged and independent of external contexts. 
For another, a user's friend who has purchased one item may recommend it to that user who then may be more likely to click such item, which is termed as \emph{social influence} \cite{Influence} (Fig.~\ref{fig-motivation}.(b)). Actually,
social influence contributes to behavior-level effect for user preference which may change dynamically with specific contexts. 
\emph{For example, a commercial analyst may be influenced by different communities of friends when it comes to an entertainment article or a commercial news.} 
Therefore, we can distinguish two components that may impact a user's preference: a static effect from social homophily and a dynamic effect from social influence\footnote{In some previous studies, there is little difference between social homophily and social influence. Differently, in this paper, we discriminate these two notions in order to better investigate the dynamic social effect.}, and we call corresponding user's preference as static and dynamic preference, respectively (left part of Fig.~\ref{fig-motivation}).

Moreover, in item domain, there exist analogous `social effects'. 
For one thing, some related items tend to possess similar attractiveness when exposed to users. 
We term this as \emph{item-to-item homophily}, and call the component of an item attribute affected by such effect as static attribute (Fig.~\ref{fig-motivation}.(c)). 
For another, if one item is popular among a certain community, then other related items would become more likely to be clicked by users in this social group. 
Such phenomenon can be termed as \emph{item-to-item influence}, which depends on specific contexts (Fig.~\ref{fig-motivation}.(d)). 
\emph{For example, a businessman and a researcher both read one article talking about the development of AI, but the businessman would prefer another article about financial information of AI companies, while the researcher would tend to click on another article about cutting-edge breakthroughs in AI.} 
We call the component of an item attribute affected by such effect as dynamic attribute. 
Finally, as shown by \textcircled{1}$\sim$\textcircled{4} in Fig.~\ref{fig-motivation}, the two-fold social effects in both user domain and item domain may jointly influence a user's decision on one item. Such phenomenon is pervasive on Epinion, Twitter, Facebook, and Netflix, etc.

\textbf{Methodologies and Results.} To these ends, in this paper, we propose \emph{DANSER (Dual graph Attention Networks for modeling multifaceted Social Effects in Recommender systems)}, a novel architecture that extends graph attention networks (GAT) \cite{GAT}. %GAT \cite{}, as a recently proposed network embedding technique, can do attentive convolutions on graph data and encode structure information into low-dimensional representations.
DANSER is composed of two dual GATs: one dual GAT for users{\textemdash}including a GAT to capture social homophily and a GAT to capture social influence{\textemdash}and the other dual GAT for items. Specifically, in user domain, one GAT to capture social homophily attentively aggregates adjacent users' embedding that reflects ones' preferences, and outputs representations for static user preference. 
Another GAT to capture social influence aims at convolution on adjacent users' context-aware preferences, which characterize relevance of users' rated items to candidate item, and outputs representations for dynamic user preference. 
%and we illustrate its effectiveness in Fig. \ref{fig-shiyitu2}. As we can see, in previous works, friends' influences to targeted user are fixed and static. The influence weight of user $u_2$ (and $u_3$) stays unchanged in both cases: i) user $u_2$ has clicked candidate item $i$, and ii) user $u_2$ has not clicked candidate item $j$. By contrast, our proposed DANSER contains two different attention weights where one stays unchanged given the friend user and the other can dynamically change under different contexts. Then the resultant force would make $u$ pay more attention to $u_1$ and $u_2$ in terms of item $i$ and to $u_1$ and $u_3$ in terms of item $j$. Such mechanism considers both global and local views for attention to friends, and improve the model expressiveness.
In item domain, two analogous GATs as a dual architecture are designed to model items' static and dynamic attributes under effects of item-to-item homophily and influence, respectively, among related items. 
Such dual mechanism possesses two advantages: i) GATs to capture homophily and influence could collaboratively model two-fold social effects, taking advantage of both global and local views to investigate user-item interactions; ii) two GATs in item domain incorporate information from related items, which could further alleviate data sparsity problem.

Then, since user static ({resp.} dynamic) preference as well as item static ({resp.} dynamic) attribute could jointly affect a user's decision on one item (shown by \textcircled{1}$\sim$\textcircled{4} in Fig.~\ref{fig-motivation}) and the importance of these four interactions would vary for distinct user-item pairs, we propose a policy-based strategy to dynamically weigh the four interactions. 
Specifically, we model the problem as a contextual multi-armed bandit \cite{Multi-armed}, and treat the weighing strategy as a policy conditional on the context (the targeted user-item pair). 
Then our goal is to optimize a reward (w.r.t predicted loss). 
For model training, we use stochastic policy gradient to update our neural network-based policy unit, and design a local-graph aware regularization technique to reduce computational cost for regularization. 
To verify our model, we conduct experiments on one benchmark datasets from \textit{Epinions} and a commercial dataset from \textit{WeChat Top Story}. The results show that DANSER outperforms state-of-the-art models under both explicit feedback settings ($2.9\%$ MAE improvement) and implicit feedback settings ($4.5\%$ AUC improvement).

\textbf{Our contributions} can be summarized as follows:

i) \textbf{General Aspects:} We distinguish the social homophily and social influence notions in view of static and dynamic effects. 
Also, we extend the two-fold social effects in user domain to item domain, and therefore investigate four social effects in recommender systems. 
These general aspects enable the model to capture more information and improve its capacity.

ii) \textbf{Novel Methodologies:} We propose DANSER with two dual GATs and a policy-based fusion unit. {We are one of the first to use GAT for social recommendation task}, and
the dual GATs can collaboratively model four social effects in both user and item domains. 
The policy unit, based on a contextual multi-armed bandit, dynamically weighs four interactions of social effects in two domains according to specific contexts. 

%iii) \textbf{Effective Techniques:} We harness stochastic policy gradient to train the policy unit, which adds random pattern into training and helps to improve model performance. We also design a new local-graph aware regularization trick to achieve an efficient regularization.

iii) \textbf{Multifaceted Experiments:} We deploy DANSER on one benchmark dataset and a commercial dataset. 
The experiment results verify the superiority of DANSER over state-of-the-art techniques, the effectiveness of the proposed components, as well as its good interpretability for modeling social effects.
\vspace{-5pt}
\section{Preliminary and Background}

In this paper, we consider a user-item interaction matrix $\mathbf R=\{r_{ui}\}_{M\times N}$, where $M$ and $N$ are numbers of users and items respectively. 
For a recommender system with \emph{implicit} feedback, $\mathbf R$ consists of 0 and 1, where $r_{ui}=1$ if user $u$ has clicked item $i$, and $r_{ui}=0$ otherwise. 
For a recommender system with \emph{explicit} feedback, $r_{ui}$ represents the rating value given by user $u$ on item $i$ (In Epinions, rating values range from 1 to 5 and user could give a high rating if she likes the item), and $r_{ui}=0$ if user $u$ has not rated item $i$. 
We use $R_I(u)$ and $R_U(i)$ to respectively denote the set of items rated by user $u$ and the set of users who have rated item $i$. 
%Moreover, \pw{we} assume $G_U=(V_U, E_U)$ to be \pw{a} trust or friendship network for users, and $V_U$, $E_U$ are \pw{the} vertex and edge sets respectively. 
Moreover, we assume the existence of a trust or friendship network for users, which is a graph $G_U=(V_U, E_U)$ where $V_U$ is the set of users and $E_U$ is the set of edges that connect two users.
We use $F_U(u)$ to denote the set of nodes adjacent to $u$ in $G_U$.
Also, we consider interaction frequencies between user $u$ and user $v$ as edge features, denoted by $\mathbf e_{uv}\in \mathbb R^{C}$, where each element in $\mathbf e_{uv}$ denotes the frequency of one-type interaction between $u$ and $v$, and $C$ is the number of interaction types. The edge features reflect link strengths between two users.

\textbf{Problem Formulation.} The social recommendation problem can be defined as \cite{review}: given observed interaction records in $\mathbf R$ and trust relationship $G_U$, we are to estimate unobserved interactions in $\mathbf R$, i.e., the probability of a targeted user $u^+$ clicking an unobserved candidate item $i^+$ (for implicit feedback) or the rating value user $u^+$ will score item $i^+$ (for explicit feedback).

\begin{figure*}[h]
	\centering
	\includegraphics[width=\textwidth,angle=0]{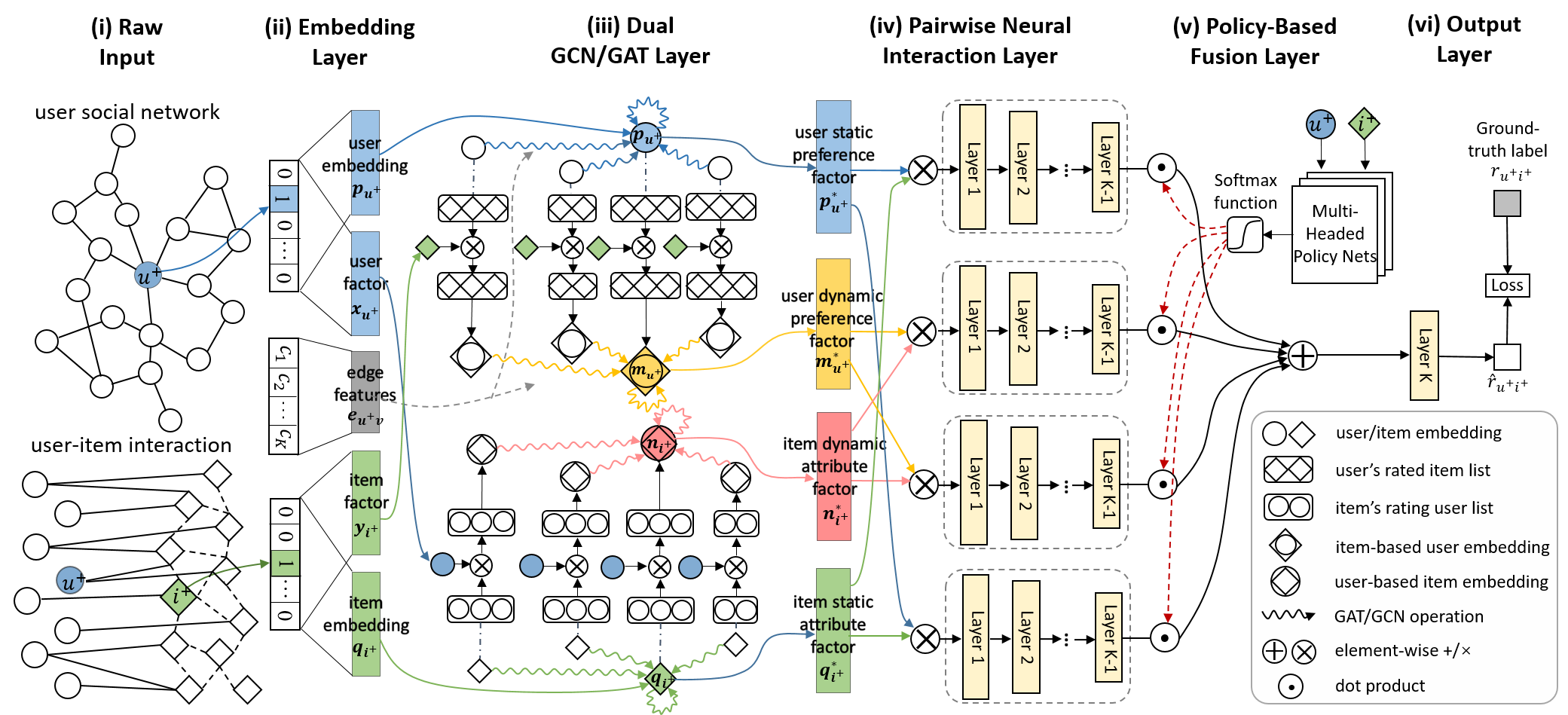}
	\caption{DANSER framework. Blue circles denote a targeted user $u^+$ while green rhombuses denote a candidate item $i^+$.  
	i) The model requires user-item interaction records and user social network as raw input. 
	We use common users who rate both items to calculate the item-item relevance and link the related items to form an item implicit network (the dotted lines between items). 
	ii) In the embedding layer, we represent one user (resp. item) as a low-dimensional embedding vector and a latent factor. Besides, interaction frequencies between users are used as edge features. iii) In the dual GCN/GAT layer, four different graph attention networks are to capture the two two-fold social effects, where the upper (resp. lower) two of them output representations for user (resp. item) static and dynamic preferences (resp. attributes) under the effect of homophily and influence, respectively. 
	iv) These four deep factors will be pairwisely combined as four interacted features, which are then fed into four independent neural networks to obtain more condensed representations. 
	v) Then a policy net with the input of item $i^+$'s and user $u^+$'s embeddings as context information outputs weights for four interacted features, which will be aggregated as one synthetic vector. 
	vi) Finally, the synthetic vector is input into the output layer to give the final predicted score $\hat r_{u^+i^+}$.}
	\label{fig-framework}
	\vspace{-10pt}
\end{figure*}

\section{Methodologies}

%In this section, we zoom in on our proposed model DANSER (Dual graph Attention Networks for modeling multifaceted Social Effects in Recommender systems).
We first introduce the model framework. For each part, we start from the motivations and intuitions, and then introduce the technical details.
Secondly, we present some training techniques to improve the model performance. 
Finally, we discuss two key units in our model for justification.

\subsection{Model Framework}

Fig.~\ref{fig-framework} provides an overview of DANSER.
We will go into the details of each part in order.

\textbf{Raw Input and Item Implicit Network}. 
The model requires user-item interaction matrix $R$ and user social network $G_U$ as input. 
Social recommendation methods focus on leveraging the user social network to solve the data sparsity problem in recommender systems. 
Most existing methods treat items independently since there is no prior information that explicitly expresses the relationship between items. 
One way to calculate the similarity or relevance between two items is by the common users who clicked or rated them \cite{item-based}. 
For any item $i$ and item $j$, we define their similarity coefficient $s_{ij}$ as the number of users who clicked both items.  
These coefficients induce an equivalence relation over items as follows: item $i$ is related to item $j$ if $s_{ij}>\tau$ with $\tau$ a fixed threshold.
We define the item implicit network as the graph $G_I=(V_I, E_I)$ where $V_I$ is the set of items and $E_I$ is the set of edges that connects two related items. 
%Then for item $i$, we define \pw{the set of its related items, denoted  $F_I(i)$, as the set of} items $j$ that satisf\pw{y} $s_{ij}>\tau$ \pw{with $\tau$ a fixed threshold.} 
%\pw{Finally,} we \sout{can} link related items and form an item implicit network $G_I=(V_I, E_I)$.

\textbf{Embedding Layer}. {The raw input of each user (resp. item) is a one-hot vector with high dimension, and the embedding operation is to project each user to a low-dimensional representation.} Inspired by \cite{SVD++} and \cite{DELF}, each user can be represented by a user-specific embedding and the items rated by her. The former representation reflects user's interests while the latter one (called \textit{item-based user embedding}) captures implicit influence of her rated history on current decision. Such embedding could deal with false-negative samples and improve model accuracy \cite{SVD++}. 
Similarly, each item can be represented by an item-specific embedding and the users who rated it (called \textit{user-based item embedding}).

For the user-specific embedding, we define $\mathbf{P}=\{\mathbf {p}_u\}_{D\times M}$, where $D$ is the embedding dimension and $\mathbf {p}_u$ denotes the embedding vector for user $u$. For the item-based user embedding, we define another embedding lookup $\mathbf {Y}=\{\mathbf {y}_i\}_{D\times N}$, where $\mathbf{y}_i$ denotes the item latent factor. Then each item in $R_I(u)$ will be mapped via $\mathbf {Y}$ and outputs $\mathcal Y_u=\{\mathbf {y}_j|j\in R_I(u)\}$, which will be used to compute the item-based user embedding, a $D$-dimensional vector representation, in the next layer. The user embeddings reflect a user's inherent preference (the `inherent' notion emphasizes that it contains no social information and only depends on the user oneself). 
Similarly, for the item-specific embedding, we define $\mathbf {Q}=\{\mathbf {q}_i\}_{D\times N}$, where $\mathbf {q}_i$ denotes the embedding vector for item $i$. For user-based item embedding, we define $\mathbf {X}=\{\mathbf {x}_u\}_{D\times M}$, where $\mathbf x_u$ is user latent factor, and each user in $R_U(i)$ will be mapped via $\mathbf {X}$ and outputs $\mathcal {X}_i=\{\mathbf {x}_v|v\in R_U(i)\}$, which will be used to compute user-based item embedding, a $D$-dimensional vector representation, in the next layer. Also, the item embeddings reflect item's inherent attribute. 

\textbf{Dual GCN/GAT Layer}. 
Graph convolution network (GCN) \cite{GCN1} conducts local convolutional operation over neighbor nodes in graph and outputs a new representation for each node, in order to encode the graph structure information as low-dimensional node representations. 
Such operation can be viewed as an extension of convolution neural network (CNN) from a grid structure to general graphs. GCN equally aggregates the neighbors' embedding in each convolution and treats each neighbor nodes with equal importance. 
In contrast, Graph Attention Network (GAT) \cite{GAT} leverages attention mechanism to consider different weights from neighbor nodes, which enables the model to filter out noises and focus on important adjacent nodes.

%In DANSER, we adopt GCN/GAT to non-linearly aggregate the embeddings of neighbor\sout{ed} nodes in both user social network $G_U$ and item implicit network $G_I$, so that the social effects from users' friends and related items can propagate to adjacent nodes through edges in graph. 
In order to exploit user social network $G_U$ and item implicit network $G_I$, we propose dual graph attention networks, an extension of GCN/GAT as a means to non-linearly aggregate the embeddings of the neighbor nodes in both networks, so that the social effects from friends and related items can propagate to adjacent nodes through the graphs.
As discussed in Section $1$, we can distinguish two types of social effects for users (resp. items) in social recommender systems: \emph{homophily} effect and \emph{influence} effect. 
They jointly affect user preferences and item attributes in different ways. DANSER uses two dual GATs to collaboratively learn different deep representations for user static/dynamic preference and item static/dynamic attribute. 

\emph{I. GAT to capture social homophily (marked as blue in Fig. \ref{fig-framework}).} 
Firstly, via user embedding, we have $\mathbf {P}$ as representation of inherent user preference factor. 
Then the GCN/GAT operation could output a new representation{, the user static preference factor $\mathbf P^*$,} by
\begin{equation}
\mathbf P^* = \sigma(\mathbf {A}_P(G_U)\mathbf {P}\mathbf {W}_P^T+\mathbf {b}_P),
\end{equation}
where $\sigma, \mathbf{{W}_P}$, $\mathbf{{b}_P}$ are activation function, weight matrix, bias vector respectively, and $\mathbf{P^*}$ is the {updated} representation of users, which incorporates social information by using the attention weights $\mathbf{A_P(G_U)}= \{\alpha^P_{uv}\}_{M\times M}$ obtained from $G_U$. 
Its elements are defined as follows:
\begin{equation}\nonumber
\alpha^P_{uv} = \frac{ attn_U(\mathbf W_P \mathbf p_u, \mathbf W_P \mathbf p_v, \mathbf W_E \mathbf e_{uv})}{\sum_{w\in  \Gamma_U(u)}attn_U(\mathbf W_P \mathbf p_u, \mathbf W_P \mathbf p_w, \mathbf W_E\mathbf e_{uv})}, v\in \Gamma_U(u),
\end{equation}
where $\mathbf W_E$ is weight matrix, $\Gamma_U(u) = \{u\}\cup F(u)$, $attn_U(x,y,z)=LeakyReLu(\mathbf w_U^Tz\otimes(x||y))$, and $\mathbf w_U$ is weight vector. Here $\otimes$ and $||$ denote the element-wise product and concatenation, respectively.
Note that the above GAT weight $\alpha^P_{uv}$ remains unchanged given user $u$ and $v$, which means factor matrix $\mathbf P^*$ is fixed for users. 
This design conforms to the intuition that social homophily contributes to static effect for user's preference, {this is why} we call it static preference factor.

\emph{II. GAT to capture social influence (marked as yellow in Fig. \ref{fig-framework}).} 
In contrast to social homophily, social influence effect is often context-aware, and the model needs to output different attention weights for friends w.r.t different candidate items. 
Through the embedding layer, we have $\mathcal Y_{u}$ as the embedding of {the items clicked by} user $u$.
Then we let each item clicked by user $u$ interact with candidate item $i^+$, 
\begin{equation}\nonumber
\mathcal Y^{i^+}_{u} = \{\mathbf y_j \otimes \mathbf y_{i^+}|j\in R_I(u)\}.
\end{equation}
This product operation can help to focus on the candidate item and model dynamic social influence under a specific context. 
{We define the item-based user embedding $\mathbf M_{i^+}=\{\mathbf m^{i^+}_u\}_{D\times M}$, which depends on candidate item $i^+$ with max pooling}
%= \pw{(}m^{i^+}_{u1}, m^{i^+}_{u2}, \cdots, m^{i^+}_{uD}\pw{)}$,}
to select the most dominating features for $D$ dimensions:
%\begin{equation}\nonumber
%\mathbf m^{i^+}_u = \pw{(}m^{i^+}_{u1}, m^{i^+}_{u2}, \cdots, m^{i^+}_{uD}\pw{)},
%\end{equation}
\begin{equation}\nonumber
m^{i^+}_{ud} = \max\limits_{j\in R_I(u)}\{y_{jd} \cdot y_{{i^+}d} \} \quad \forall d=1,\ldots, D
\end{equation}
   {where} %$\mathbf m^{i^+}_u$ is \pw{the} item-based user embedding for user $u$, and 
$m^{i^+}_{ud}$, $y_{{i^+}d}$, {and} $y_{jd}$ are the $d$-th feature of $\mathbf m^{i^+}_u$, $\mathbf y_{i^+}$, $\mathbf y_j$ respectively. 
The max pooling operation can help to focus on the most important value and alleviate noises in users' clicked history. {The item-based embedding $\mathbf m^{i^+}_u$ includes two information}: i) user $u$'s context-aware preference (w.r.t candidate item $i^+$), and ii) inherent representation for $u$ (independent of social information). 
{In order to define the user dynamic preference factor $\mathbf M_{i^+}^*$,}
we proceed to incorporate the social information from friend users,
\begin{equation}\nonumber
\mathbf M_{i^+}^* = \sigma(\mathbf A_M(G_U)\mathbf M\mathbf W_M^T+\mathbf b_M),\mathbf A_M(G_U) = \{\alpha^M_{uv,{i^+}}\}_{M\times M},
\end{equation}
\begin{equation}\nonumber
\alpha^M_{uv,{i^+}} = \frac{ attn_U(\mathbf W_M \mathbf m^{i^+}_u, \mathbf W_M \mathbf m^{i^+}_v, \mathbf W_E\mathbf e_{uv})}{\sum_{w\in  \Gamma_U(u)}attn_U(\mathbf W_M \mathbf m^{i^+}_u, \mathbf W_M \mathbf m^{i^+}_w, \mathbf W_E\mathbf e_{uv})},
\end{equation}
for $v\in \Gamma_U(u)$.
Note that the above attention weight $\alpha^M_{uv,i^+}$ depends on the user's history of rated items as well as specific candidate item $i^+$, which indicates that factor matrix $\mathbf M_{i^+}^*$ would change dynamically with  different contexts. This design conforms to the intuition that social influence contributes to context-aware effect for a user's preference, so we term it as dynamic preference factor.

\emph{III. GAT to capture item-to-item homophily (marked as green in Fig. \ref{fig-framework}).} Similarly, we use the item embedding $\mathbf Q$ as representation of inherent item attribute factor, and then leverage GAT to incorporate social information,
\begin{equation}\nonumber
\mathbf Q^* = \sigma(\mathbf A_Q(G_I)\mathbf Q\mathbf W_Q^T+\mathbf b_Q), \mathbf A_Q(G_I) = \{\alpha^Q_{ij}\}_{N\times N},
\end{equation}
\begin{equation}\nonumber
\alpha^Q_{ij} = \frac{ attn_I(\mathbf W_Q \mathbf q_i, \mathbf W_Q \mathbf q_j)}{\sum_{k\in  \Gamma_I(i)}attn_I(\mathbf W_Q \mathbf q_i, \mathbf W_Q \mathbf q_k)}, j\in \Gamma_I(i),
\end{equation}
where $attn_I(x,y) = LeakyRelu(\mathbf w_I^T(x||y))$, and $\mathbf w_I$ is a weight vector. The GAT weight $\alpha^Q_{ij}$ remains unchanged given item $i$ and $j$. Correspondingly, item-to-item homophily contributes to static effect for item's attribute, and we call it static attribute factor.

IV. \emph{GAT to capture item-to-item influence (marked as red in Fig. \ref{fig-framework}).} We now model item-to-item influence, which is context-aware and {depends on the} specific targeted user. Hence, our model needs to output different attention weights for distinct related items w.r.t different targeted users. 
{Similarly to social influence, the user-based item representation $N_{u^+} = (n^{u^+}_i)_{D\times N}$ for a given targeted user $u^+$  can be defined as follows: 
\begin{equation}\nonumber
\mathcal X^{u^+}_i = \{\mathbf x_v \otimes \mathbf x_{u^+}|v\in R_I(i)\},
\end{equation}
\begin{equation}\nonumber
n^{u^+}_{id} = \max\limits_{v\in R_U(i)} \{x_{vd}\cdot x_{{u^+}d}\}\quad \forall d=1, \ldots, D,
\end{equation}
where $n^{u^+}_{id}$, $x_{{u^+}d}$, {and} $x_{vd}$ are the $d$-th feature of $\mathbf n^{u^+}_i$, $\mathbf x_{u^+}$, $\mathbf x_v$ respectively.
Then} the representation for item dynamic attribute $\mathbf N_{u^+}^*$ can be computed as follows:
\begin{equation}\nonumber
\mathbf N_{u^+}^* = \sigma(\mathbf A_N(G_U)\mathbf N_{u^+}\mathbf W_N^T+\mathbf b_N),\mathbf A_N(G_U) = \{\alpha^N_{ij,{u^+}}\}_{N\times N},
\end{equation}
\begin{equation}\nonumber
\alpha^N_{ij,{u^+}} = \frac{ attn(\mathbf W_N \mathbf n^{u^+}_i, \mathbf W_N \mathbf n^{u^+}_j)}{\sum_{k\in  \Gamma_I(i)}attn(\mathbf W_N \mathbf n^{u^+}_i, \mathbf W_N \mathbf n^{u^+}_k)}, j\in \Gamma_I(i).
\end{equation}

\textbf{Pairwise Neural Interaction Layer}. Since user's decision on one item often depends on both user preference and item attribute, we can let two preference factors and two attribute factors pairwisely interact with each other. Then we borrow the idea from \cite{NCF}, feeding the four results into different neural networks indexed by $a\in \{1,2,3,4\}$:
\begin{equation}\nonumber
\mathbf s_a = \phi_K^a(\cdots\phi_2^a(\phi_1^a(\mathbf z_0[a]))),
\end{equation}
\begin{equation}\nonumber
\phi_k^a(\mathbf z_{k-1}) = tanh(\mathbf W_k^a \mathbf z_{k-1}^a+\mathbf b_k^a), k\in[1, K-1],
\end{equation}
\begin{equation}\nonumber
\mathbf z_0 = [\mathbf p^*_u \oplus \mathbf q^*_i, \mathbf p^*_u \oplus \mathbf n^*_i, \mathbf m^*_u \oplus \mathbf q^*_i, \mathbf m^*_u \oplus \mathbf n^*_i],
\end{equation}
%where $a\in \{1,2,3,4\}$. 
Here we employ
a tower structure for each network, where higher layers have
smaller number of neurons \cite{NCF}. 

\textbf{Policy-Based Fusion Layer}. 
We further fuse the four interacted features $\mathbf s_a$ into a synthetic one. Note that homophily effect and influence effect could jointly affect user preference and item attribute, but for distinct users and items, the importance of the two-fold social effects could be different. 
To this end, we propose a new policy-based fusion strategy to dynamically allocate  weights to the four interacted features according to specific user-item pairs.

We model the weight allocation as a contextual multi-armed bandit problem\cite{Multi-armed},where an action, denoted by $\gamma\in\{1, 2, 3, 4\}$, indicates which feature to choose, a context is a user-item pair and the reward after playing an action represents a recommendation loss. In this problem, a stochastic policy can be written as the conditional probability $p(\gamma|\mathbf p_u, \mathbf q_i)$, which could characterize the selecting probability or importance weights for different combinations of social effects given a specific pair $(u,i)$.

To make the problem solvable, we approximate $p(\gamma|\mathbf p_u, \mathbf q_i)$ by a neural network (called policy network):
\begin{equation}\nonumber
e_\gamma = \mathbf W^2_F\tanh(\mathbf W^1_F (\mathbf p_u || \mathbf q_i)+\mathbf b^1_F)+\mathbf b^2_F.
\end{equation}
%\begin{equation}\nonumber
%\mathbf g_0 = [\mathbf p_u, \mathbf q_i]
%\end{equation}
\begin{equation}\nonumber
p(\gamma|\mathbf p_u, \mathbf q_i) = \frac{\exp(e_\gamma)}{\sum_{a=1}^4 \exp(e_a)}.
\end{equation}
Then the synthetic representation can be expressed as
\begin{equation}\nonumber
\mathbf s = \mathbb E_{\gamma\sim p(\gamma|\mathbf p_u, \mathbf q_i)}(\mathbf s_\gamma)=\sum\limits_{\gamma=1}^4 p(\gamma|\mathbf p_u, \mathbf q_i)\cdot \mathbf s_\gamma.
\end{equation}
We call the above strategy {\it single-headed} policy-based fusion. 
Inspired by \cite{AIAYN}, we can extend it to a {\it multi-headed} version. 
We harness $L$ different independent policy networks and the final weights are given by the averaged weights given by each policy net. The training of policy networks is by stochastic policy gradient, which will be discussed in Section 3.2.

\textbf{Output Layer}. Then probability of user $u$ clicking item $i$ can be predicted by
\begin{equation}\nonumber
\hat r_{ui} = nn(\mathbf s).
\end{equation}
If a clicking probability is required (for implicit feedback), $nn(\cdot)$ can be a fully-connected layer with a sigmoid activation function. 
If a rating value is needed (for explicit feedback), $nn(\cdot)$ can be a fully-connected layer without activation function. 

\textbf{Loss Function}. The loss function measures the discrepancy between predicted value and ground-truth value. For implicit feedback, the most widely adopted loss function is the cross-entropy defined as
\begin{equation}\nonumber
\mathcal L_1=-\sum\limits_{(u,i)}r_{ui}\log \hat r_{ui}+(1-r_{ui})\log (1-\hat r_{ui}).
\end{equation}
For explicit feedback, we adopt the mean square loss
\begin{equation}\nonumber
\mathcal L_1=\sum\limits_{(u,i)}\|\hat r_{ui}-r_{ui}\|^2.
\end{equation}
Here, we concentrate on pointwise loss. Also, other pairwise loss functions can be used and we leave them for future investigation.

\subsection{Training}
In this subsection, we introduce some training techniques to improve the model performance.

\textbf{Mini-Batch Training}. We also leverage mini-batch training to calculate the gradient. For each iteration, we consider $B$ user-item pairs. For each pair $(u,i)$, the neighbor nodes $F_U(u)$, $F_I(i)$ in two networks as well as the rated history $R_I(v)$, $v\in\Gamma_U(u)$, and $R_U(j)$, $j\in\Gamma_I(i)$ will be together input into the model to compute the gradient. Once an epoch is finished, the batch partition will be randomly reset to incorporate enough noise during optimization. In each mini-batch, since numbers of friends are different for users, we need to pad zeros and input a matrix with dimension of max friend number in one batch. We observe that number of friends tends to obey a long-tail distribution, so the padding would lead to both high space and time complexity. Therefore, we use a sample technique: 1) for friend number more than $F$, we uniformly sample $F$ friends as input data; 2) for friend number less than $F$, we pad zeros to obtain a $F$-dimensional vector. We also experiment other sample methods like random walk with restart (RWR) \cite{RWR} on a sub-graph near central node, but we find that uniformly sampling over adjacent nodes can achieve competitive performance. This observation could be justified by one argument in sociology research that influence from users always propagates to adjacent users in social networks \cite{WSDM}. Thus we adopt the simpler method to reduce time complexity. In experiment part, we will investigate the impact of sample size $F$ on the performance.

\textbf{Local-Graph Aware Regularization}. The other issue is regularization to avoid over-fitting problem. In our model, we adopt L1 regularization to constrain the embedding parameters to sparse forms. The regularization loss can be expressed as
\begin{equation}\nonumber
\mathcal L_2 = \sum\limits_u (\|\boldmath p_u\| + \|\boldmath x_u\|) + \sum\limits_i(\|\boldmath q_i\| + \|\boldmath y_i\|).
\end{equation}
However, the above loss introduces high computational complexity, since for each batch training we need to calculate the gradient of regularization over all users and items. Such operation can not generalize to large-scale data set. A recent work \cite{deepinterest} proposes a mini-batch aware regularization trick to solve this problem. In each batch training, only the embedding parameters for user-item pair $(u,i)$ in the batch will be used to calculate the regularization loss. Here we extend this idea to graph structure and call it as local-graph aware regularization. In each batch training, we constrain the embedding parameters of users, items as well as their neighbor nodes in social network and item implicit network. The new regularization loss can be written as
\begin{eqnarray}\nonumber
\begin{aligned}
\mathcal L_2 &=\frac{1}{2}\sum\limits_{(u,i)} \big[\|\mathbf p_u\| + \|\mathbf x_u\|+
\sum\limits_{v\in \Gamma_U(u)} \frac{1}{|F_U(v)|}(\|\mathbf p_v\| + \|\mathbf x_v\|) \\
&+\|\mathbf q_i\| + \|\mathbf y_i\| + \sum\limits_{j\in \Gamma_I(i)} \frac{1}{|F_I(j)|}(\|\mathbf q_j\| + \|\mathbf y_j\|)\big].
\end{aligned}
\end{eqnarray}
In fact, the local-graph aware regularization is to split the regularization loss into small parts for each mini-batch and guarantees that once an epoch is finished, each user and each item can be regularized equally.
To sum up, the final loss function is
\begin{equation}\label{eq-loss}
\mathcal L = \mathcal L_1 + \lambda \mathcal L_2,
\end{equation}
where $\lambda$ is a trade-off parameter between accuracy and complexity. 

Besides, for parameters in neural networks (weight matrix and bias vectors), we adopt dropout strategy to replace the traditional regularization. Our experiments will further discuss the influence of different dropout probabilities $\rho$.

\textbf{Policy Gradient.} In training stage, we train the policy networks in a stochastic way. Assume $p_l(\gamma|\mathbf p_u, \mathbf q_i)$ to be the active probability given by $l$-th policy network, and random variable $\gamma$ obeys a multinomial distribution. For each policy network, we draw $\gamma\sim Multi(p_l(\gamma|\mathbf p_u, \mathbf q_i))$, and feed $\mathbf s_\gamma$ into output layer, which gives prediction and returns loss. We leverage the loss to train the policy network. This design can be formulated as an actor-critic terminology: i) given the environment (user $u$ and item $i$), the $l$-th agent (policy net) needs to execute an action according to policy ($p_l(\gamma|\mathbf p_u, \mathbf q_i)$); ii) every action would generate a reward (recommendation loss) that could lead the training of policy network. We define the reward as subtraction of loss generated by selected interacted feature, i.e., $\mathcal R(\boldmath p_u, \boldmath q_i, \gamma)=-\mathcal L(\mathbf s_\gamma)$ (where $\mathcal L(\mathbf s_\gamma)$ denotes the loss when we input $\mathbf s_\gamma$ into the output layer). We aim at maximizing the expected reward $\mathbb E_{\gamma\sim p(\gamma|\boldmath p_u, \boldmath q_i)}(\mathcal R(\boldmath p_u, \boldmath q_i, \gamma))$. The policy gradient method REINFORCE \cite{REIN} can be used to update parameters in one policy network (denoted as $\theta$), and the gradient can be derived as follows:
\begin{eqnarray}
\begin{aligned}\label{eq-policy}
&\nabla_\theta \mathbb E_{\gamma\sim p_\theta(\gamma|\mathbf p_u, \mathbf q_i)}(\mathcal R(\mathbf p_u, \mathbf q_i, \gamma))\\
%=&\sum\limits_\gamma \nabla_\theta p_\theta(\gamma|\mathbf p_u, \mathbf q_i)\mathcal R(\mathbf p_u, \mathbf q_i, \gamma)\\
%=&\sum\limits_\gamma  p_\theta(\gamma|\mathbf p_u, \mathbf q_i) \nabla_\theta\log p_\theta(\gamma|\mathbf p_u, \mathbf q_i)\mathcal R(\mathbf p_u, \mathbf q_i, \gamma)\\
%=&\mathbb E_{\gamma\sim p_\theta(\gamma|\mathbf p_u, \mathbf q_i)}[\nabla_\theta\log p_\theta(\gamma|\mathbf p_u, \mathbf q_i)\mathcal R(\mathbf p_u, \mathbf q_i, \gamma)]\\
\simeq& \frac{1}{4}\sum\limits_\gamma\nabla_\theta\log p_\theta(\gamma|\mathbf p_u, \mathbf q_i)\mathcal R(\mathbf p_u, \mathbf q_i, \gamma).
\end{aligned}
\end{eqnarray}

\textbf{Training Algorithm.} The training of DANSER updates policy networks and feedforward networks (including embedding, dual GAT, pairwise neural interaction and output layers) iteratively, which is presented in Alg.~\ref{alg}. First, we update feedforward networks for $n_p$ steps (one mini-batch for one step) by optimizing \eqref{eq-loss}. Then we train $L$ policy networks independently by conducting corresponding policy gradient \eqref{eq-policy}.

\begin{algorithm}[h]
	\caption{Training for DANSER}
	\label{alg}
	\textbf{REQUIRE:} $\mathbf R$, observed user-item interaction. $G_U$, user social network. $G_I$, item implicit network.\\
	\textbf{REQUIRE:} $\eta$, $\zeta$, learning rates.\\
	\While{not converged}
	{\For{$i=1,\cdots,n_p$}{
			Sample $B$ user-item pairs.\;
			For each pair $(u,i)$, uniformly sample $F$ neighbors from $F_U(u)$ and $F_I(i)$, respectively\;
			Draw $\gamma_l\sim Multi(p^l_{\theta_l}|\mathbf p_u, \mathbf q_i)$, $l=1,\cdots,L$\;
			$w\leftarrow \eta\nabla_w \mathcal \sum_{l}L(\mathbf s_{\gamma_l})$\;}
	$\theta_l \leftarrow \zeta \nabla_{\theta_l} \mathbb E_{\gamma\sim p^l_{\theta_l}(\gamma|\mathbf p_u, \mathbf q_i)}(\mathcal R(\mathbf p_u, \mathbf q_i, \gamma))$, $l=1,\cdots,L$\;}
\end{algorithm}

\textbf{Complexity Analysis.} Since complexity and scalability are two important concerns for graph algorithm, we investigate complexity of DANSER for model training and inference respectively. For training, we need to update embedding of all friend users and related items for $B$ user-item pairs in one batch, which requires time complexity $O(BFCD)$, where $C=\max\{|R_I(v)||v\in \Gamma_U(u)\}\cup\{|R_U(j)||j\in \Gamma_I(i)\}$. We can control $C$ to a limited value by truncating recent $C_t$ rated items for users. The friend number $F$ is controlled by our sample method. Also, for inference, we need $O(BFCD)$ to compute GAT representations. Hence, DANSER can scale linearly w.r.t number of user-item interactions.

\subsection{Discussions}

In this subsection, we discuss on effectiveness of our key units, dual GAT layer and policy-based fusion layer, in order to shed more insights on DANSER.

\textbf{Justification of Dual GATs.}
{During the} training process, the GAT to capture {the} homophily effect would give larger weights to friends whose preferences for items are globally similar to {the} targeted user (resp. related items whose attractiveness to users are globally similar to candidate item), so the static preference (resp. attribute) factor captures static effect of friend users (resp. related items) from a global view. Differently, the weights given by the GAT to capture {the} influence effect depend on {the} targeted user (resp. item), so the corresponding dynamic factor captures context-aware effect{s} from a local view. In each domain, a dual GAT collaboratively models {the} two-fold effects and takes both global and local views to conduct a more accurate investigations into distinct users and items.

\begin{figure}[t]
\vspace{-5pt}
\setlength{\abovecaptionskip}{-0.1cm}
	\centering
	\includegraphics[width=0.47\textwidth,angle=0]{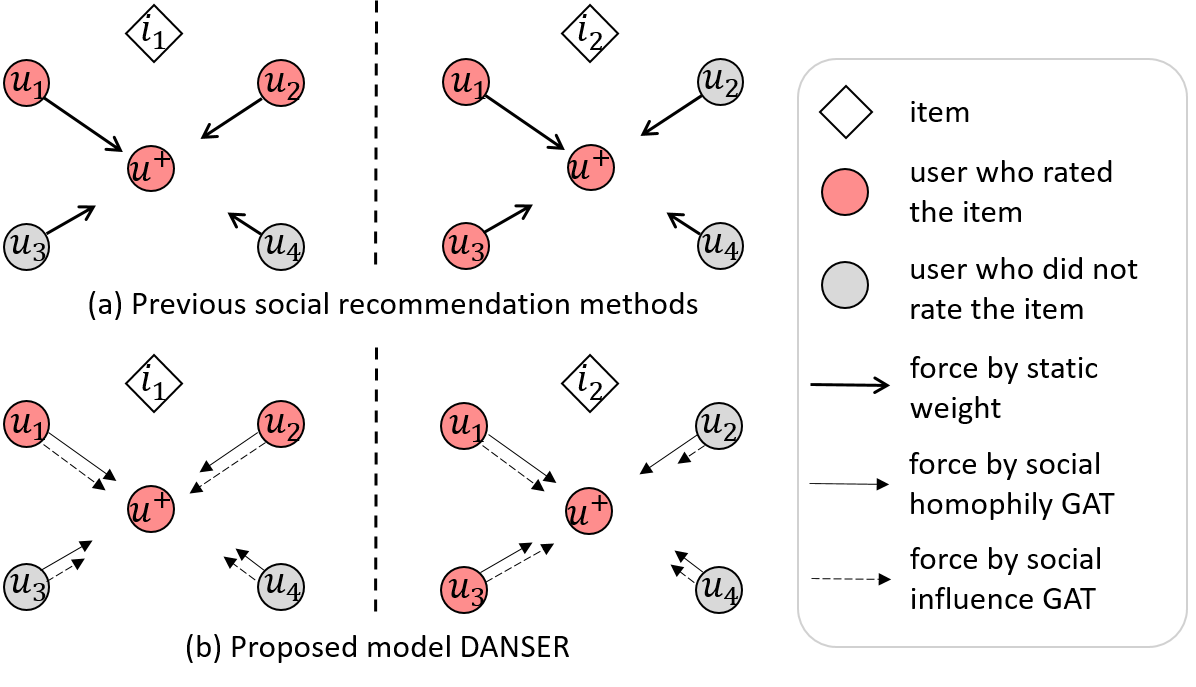}
	\vspace{2pt}
	\caption{Illustration of effectiveness of dual GATs and comparison with previous social recommendation methods.}
	\label{fig-shiyitu2}
	\vspace{-10pt}
\end{figure}

{
To put it more symbolically, we illustrate its effectiveness by  considering the following two cases for the recommendation of candidate item $i^+$ to targeted user $u^+$: item $i^+$ is similar to item $i_1$ and $i^+$ is similar to $i_2$ (see left and right columns of Fig. \ref{fig-shiyitu2}). 
Here, we assume that user $u^+$ has only clicked on $i_1$ and $i_2$ (and no other items), and 
users $u_1$ {to} $u_4$ are $u^+$'s friends. 
%As seen on Fig. \ref{fig-shiyitu2}, $u_1$ has also clicked on both items $i_1$ and $i_2$, whereas $u_2$ and $u_3$ has clicked only on one of the two items. 
%User $u_4$ has clicked on none of them.
}
%\qt{Assume that $u_1$'s preference is quite similar to $u$, $u_2$ and $u_3$ are partially similar to $u^+$, and $u_4$ is not similar to $u^+$}.
Previous works for social recommendation consider static effect of friends in an average manner. 
{As {illustrated} in Fig. \ref{fig-shiyitu2}.(a), for friend $u_1$ who {clicked on} both items, the attention weight would be large, and {the opposite} for friend $u_4$. 
For $u_2$ and $u_3$, their weights are medium since they are both partially similar to $u^+$. }
{Therefore} static weights cannot discriminate {between the} influence of user $u_2$ and $u_3$ when it comes to item {$i^+$ even if it is known that it is similar to $i_1$ or $i_2$}. 
{In} contrast, DANSER contains two different attention weights, where one captures user-specific static effect (for homophily) and the other can dynamically change under different contexts (i.e., for influence).
{When candidate item $i^+$ is similar to $i_1$ (resp. $i_2$), the social influence GAT would focus more on $u_1$ and $u_2$ (resp. $u_1$ and $u_3$).}
%Since $u^+$ is similar to $u_1$ and $u_2$ (resp. $u_1$ and $u_3$) for item $i^+_1$ (resp. $i^+_2$), the social influence GAT would focus more on $u_1$ and $u_2$ (resp. $u_1$ and $u_3$) when the candidate item is $i^+_1$ (resp. $i^+_2$). 
Then the resultant force of {the} two GATs would make $u^+$ pay more attention to $u_2$ %in terms of item $i^+_1$ 
{when $i^+$ is similar to $i_1$}
and to $u_3$ %in terms of item $i^+_2$. 
{in the other case.}
The global and local views of DANSER work collaboratively for selecting {the} `right' neighbors that are similar to {the} targeted user under a specific context, which improves model expressiveness. 
In Section 4.5, we will verify the above argument via a real-case study.

\textbf{Justification of Policy-Based Fusion.}
The policy-based fusion layer acts like a stochastic network while training, and the REINFORCE algorithm can help to explore the optimal weights distribution over four interacted features. Since the interacted feature with larger active probability is more likely to be selected, it would be more frequent for the interaction neural layer that gives the `best' feature to be trained. Such mechanism brings up two major benefits. Firstly, the policy networks can filter `right' user-item samples for training of corresponding interaction neural layer as well as previous GCN/GAT and embedding layers. Second, the active probability introduces random pattern to the training process, which can help to escape from local optimums. Our ablation study in experiment part also validate the superiority of the policy-based fusion compared with other straightforward strategies.

\section{Experiments}

To comprehensively evaluate proposed model DANSER\footnote{The codes are released at https://github.com/echo740/DANSER-WWW-19.}, we conduct experiments to answer the following research questions:

\noindent\textbf{RQ1} How does DANSER perform compared with state-of-the-art models for recommendation and social recommendation?

\noindent\textbf{RQ2} Are the key components in DANSER, such as dual GATs and policy-based fusion, necessary for improving performance?

\noindent\textbf{RQ3} How do hyper-parameters in DANSER impact recommendation performance?

\noindent\textbf{RQ4} How can DANSER interpret {the} four different social effects from real-world data?

\subsection{Experiment Setup}
\subsubsection{Data sets} We apply our model to one public benchmark datasets for social recommendation \emph{Epinions} and a commercial dataset \emph{WeChat Top Story}. 
We provide some statistics of these two datasets in Table \ref{table-dataset}. 
The basic information about the two datasets is summarized as follows:

i) \emph{Epinions:} Epinions is a consumer review websites, where users can rate some items and add other users in their trust lists. \emph{Epinions} dataset \cite{epinion} contains two parts of information: user-item interaction pairs, where items are rated from 1 to 5 (explicit feedback), as well as the directed trust relationships between users. The dataset has been widely used as benchmarks for social recommendation. Following other works \cite{TrustSVD,TrustMF,NSCR,SREPS}, we randomly choose $80\%$ user-item interactions as training set and the remaining $20\%$ as test set for each dataset.

ii) \emph{WeChat:} 
%WeChat is a Chinese multi-purpose messaging, social media app. It has been the world's largest standalone mobile app with over 1 billion monthly active users. Users can not only chat with their friends online through WeChat, but also share their experiences, moods and articles they love through WeChat Moments. Recently, WeChat launched a new application named Top Stories, which aims at recommending articles of potential interests to targeted users. 
We also deploy our model on a real-world article recommender system, WeChat Top Story. This dataset is constructed by user-article clicking records on this platform. Different from \emph{Epinions}, this dataset only contains implicit feedback, i.e., we only know whether a user clicked an article or not. Besides, the friend relationships are bidirectional. We chronologically order the user-article clicking records, and use the first $90\%$ records to train our model and the remaining records to evaluate the prediction performance. Since the positive samples and negative samples are fairly unbalanced in this dataset, we uniformly sample the negative ones such that the numbers of positive and negative samples are the same for one user.

\begin{table}[h]
\setlength{\abovecaptionskip}{-0.5pt}
\setlength{\belowcaptionskip}{-5pt}
	\centering
	\small
	\caption{Statistics of three datasets.}
    \label{table-dataset}
	\begin{tabular}{ccccc}
		\toprule
		Dataset & \#users & \#items & \#interactions &\#relationships \\
		\midrule
		Epinions &49,290  & 139,738  & 664,824 &487,181    \\
		WeChat &$\sim$200,000  &$\sim$100,000  & $\sim$4,000,000 & $\sim$2,000,000   \\
		\bottomrule
	\end{tabular}
	\vspace{-15pt}
\end{table}

\subsubsection{Implementation Details}

We use Tensorflow to implement our model and deploy it on GTX 1080 GPU with 8G memory. The hyper-parameter settings for \emph{Epinions} are as follows: batch size $B=64$, embedding dimension $D=10$, dropout ratio $\tau=0.5$, regularization coefficient $\lambda=0.001$, sample size $F=30$, truncation length $C_t=30$, multi-head number $L=4$, policy gradient period $n_p=1000$. The learning rates $\eta=0.1$, $\zeta=0.01$. The Leaky ReLu slope is 0.2. In pairwise interaction neural layer, we adopt three-layer neural networks and neuron numbers are $10-16-8-4$.

\subsubsection{Competitors}

We choose several comparative methods, including some state-of-the-art models for recommendation and social recommendation, to evaluate the performance of our DANSER. The optimal hyper-parameter settings for each method are determined either by our experiments or
suggested by previous works.

The recommendation models include \emph{SVD++}\cite{SVD++} and \emph{DELF}\cite{DELF}. These two methods only leverage user-item interactions as model input. They can be used to evaluate the effectiveness of other models that use social information. 
\begin{itemize}
    \item \emph{SVD++} \cite{SVD++} is a basic model-based recommendation method which use both user-specific and item-based user embedding to represent user's preferences.
    \item \emph{DELF} \cite{DELF} is a state-of-the-art CF method which proposes dual embedding for users and items, and adopts deep neural networks to capture complex non-linear information.
\end{itemize}
 
The social recommendation models include \emph{TrustPro}, \emph{TrustMF}, \emph{TrustSVD}, \emph{NSCR}, \emph{SREPS}. These methods all leverage both user-item interactions and user friendships as input information.

\begin{itemize}
    \item \emph{TrustPro} \cite{geographical} is a trust propagation method, using rating of friends to deduce rating of targeted user.
    \item \emph{TrustMF} \cite{TrustMF}, as one matrix factorization method, optimizes user embedding to retrieve the trust matrix.
    \item \emph{TrustSVD} \cite{TrustSVD} is another matrix factorization-based method, incorporating friends' embedding vectors into targeted user's predicted rating.
    \item \emph{NSCR} \cite{NSCR}, as a strong baseline for social recommendation, adopts deep neural networks to learn latent representations of users and items, and leverages graph regularization to constrain the embedding of adjacent users to be similar.
    \item \emph{SREPS} \cite{SREPS} is another strong baseline, using a network embedding approach to encode social network.
\end{itemize}
 
We also construct several variants of DANSER as ablation study. To evaluate the necessity of dual GAT layer, we simplify our model as \emph{DualEMB}, \emph{DualGCN}, \emph{userGAT}, and \emph{itemGAT}. \emph{DualEMB} removes the convolution over adjacent nodes, while \emph{DualGCN} replace GAT by GCN. \emph{userGAT} removes two GATs in item domain, and \emph{itemGAT} removes two GATs in user domain. Moreover, to evaluate the necessity of policy-based fusion layer, we consider the following variants. \emph{DANSER-m} and \emph{DANSER-a} leverage max and average pooling to fuse four interacted features, respectively. \emph{DANSER-c} concatenates the four interacted features as the input of output layer. \emph{DANSER-w} removes the policy net, set four sharing fusion weights, and update the weights together with training of feed-forward networks.

\subsubsection{Evaluation Protocol}

We adopt different metrics to evaluate recommendation performance. Since two datasets possess different feedbacks, we consider different metrics for them. For \emph{Epinions} with explicit feedback, we use \emph{MAE} and \emph{RMSE}, which are widely adopted by other works, as evaluation metrics. For \emph{WeChat} with implicit feedback, we use \emph{Precision@k} (short as P@k) and \emph{AUC}, two universally acknowledged metrics for 0-1 classification, to evaluate the performance. We repeated each experiment ten times and report the
average performance.

\begin{comment}
The definitions of these metrics are as follows:

\emph{MAE:} Mean Absolute Error (MAE) measures the overall $L_1$ distance between predicted ratings and ground-truth values:
\begin{equation}
MAE =\sum_{{(u,i)} \in \mathcal R} \frac{|\hat{r}_{u,i}-r_{u,i}|}{N}
\end{equation}
where $N$ denotes the total number of rating records, $\hat{r}_{u,i}$ and $r_{u,i}$ denote the predicted and ground-truth ratings for user $u$ on item $i$, respectively. 

\emph{RMSE:} Root Mean Square Error (RMSE) measures the overall $L_2$ distance between predicted ratings and ground-truth values:
\begin{equation}
RMSE = \sqrt{\frac{\sum_{{u,i}\in \mathcal R}(\hat{r}_{u,i}-r_{u,i})^2}{N}}
\end{equation}

\emph{AUC:} Area Under Curve (AUC) refers to the area under the ROC curve, and measures the probability that a recommender system ranks a positive user-item interaction higher than negative ones:
\begin{equation}
AUC = \frac{\sum\limits_{{i} \in \mathcal I_u^{+}} \sum\limits_{{j} \in \mathcal I_u^{-}}-\delta(\hat{r}_{uij}>0)}{|\mathcal I_u^{+}||\mathcal I_u^{-}|}
\end{equation}
where $\mathcal I_u^{+} =\{i|r_{ui}=1\}$ , $\mathcal I_u^{-} =\{i|r_{ui}=0\}$ are the sets of observed positive item $i$ and unobserved item $j$ for user u respectively, and $\delta$ is the count function returning $1$ if $\hat{r}_{uij}>0$ and 0 otherwise. 
\end{comment}

\subsection{Comparative Results: RQ1}

We report experiment results of DANSER and other comparative methods in Table \ref{tbl-result1}. As we can see, our model DANSER outperforms other comparative methods and achieves great improvements, i.e., $2.87\%$ MAE impv. for Epinions and $4.48\%$ AUC impv. for WeChat. The results verify the superiority of proposed model for social recommendation. Also, there are some findings in these comparative experiments. First, matrix factorization methods like TrustMF and TrustSVD can achieve good accuracy in Epinions with explicit feedback, but perform poorly in WeChat with implicit feedback. Second, NSCR performs quite better than other linear methods, which indicates that neural network plays an important role in capturing complex non-linear information. Third, even though DELF has not leveraged any social information, it outperforms some social recommendation methods like TrustSVD and NSCR. This is possibly due to the facts: i) the dual embedding and neural interaction layer in DELF can better exploit information in user-item interactions; ii) other social recommendation methods like TrustSVD and NSCR do not make good use of social information. Indeed, TrustSVD assumes equal importance of each friend user while NSCR uses graph regularization that constrains representations for adjacent users in a static way. These settings resort to a prior assumption that friend users' preferences are all similar, which may not be true for contemporary SNS (also discussed in Section 1).

\begin{table}[t]
	\centering
	\caption{Comparative results for Epinions and WeChat. For MAE, RMSE, the smaller value is better, and vice versa for P@10, AUC.}
	\vspace{-8pt}
	\label{tbl-result1}
	\normalsize
	\begin{threeparttable}
		\begin{tabular}{c|c|c|c|c}
			\toprule[1pt]
			\specialrule{0em}{1pt}{1pt}
			\centering  & \multicolumn{2}{c|}{Epinions}& \multicolumn{2}{c}{WeChat} \\
			\specialrule{0em}{1pt}{1pt} \cline{2-5}
			\specialrule{0em}{1pt}{1pt}
			~ & MAE & RMSE & P@10 & AUC   \\ 
			\specialrule{0em}{1pt}{1pt} \hline \specialrule{0em}{1pt}{1pt}
			SVD++ \cite{SVD++}  & 0.8321 & 1.0772 & 0.0653 & 0.7304 \\
			DELF \cite{DELF}  & 0.8115 & 1.0561 & \underline{0.0752} &\underline{0.7818}    \\ 
			\specialrule{0em}{1pt}{1pt} \midrule[0.2pt] \specialrule{0em}{1pt}{1pt}
			TrustPro \cite{geographical}  & 0.9130 & 1.1124 & 0.0561 & 0.6482  \\ 
			TrustMF \cite{TrustMF}  & 0.8214 & 1.0715 & 0.0625& 0.7005 \\ 
			TrustSVD \cite{TrustSVD} & 0.8144 & 1.0492 & 0.0664& 0.7325   \\
			NSCR \cite{NSCR}     & 0.8044 & 1.0425& 0.0736 & 0.7727  \\ 
			SREPS \cite{SREPS}  & \underline{0.8014} & \underline{1.0393} & 0.0725& 0.7745   \\
			\specialrule{0em}{1pt}{1pt} \midrule[0.2pt] \specialrule{0em}{1pt}{1pt}
			DANSER & \textbf{0.7781} & \textbf{1.0268}& \textbf{0.0823} & \textbf{0.8165}    \\
			\midrule[0.5pt]
			Impv.\tnote{1} &2.87\% &1.25\% & 9.33\% & 4.48\%\\ 
			\specialrule{0em}{1pt}{1pt}
			\bottomrule[1pt]
		\end{tabular}
		\begin{tablenotes}
	\small
	\item[1] The improvement compares DANSER with the best competitor (underlined).
\end{tablenotes}
	\end{threeparttable}
	\vspace{-10pt}
\end{table}

\subsection{Ablation Study: RQ2}

In order to verify the effectiveness of some components in our model, we do some ablation studies and the results are shown in Table \ref{tbl-result2}. We first compare DANSER with DualEMB and DualGCN. As we can see, GCNs can capture enough social information hidden in graph structure and improve recommendation accuracy (about $1.0\%$ impv. for MAE over DualEMB), while GATs can further improve performance by differentiate importance weights for friend users (about extra $0.8\%$ impv. for MAE over DualGCN). We secondly compare DANSER with userGAT and itemGAT. The results show that incorporation of user and item networks can both bring significant performance improvement, and the benefits brought by user social network are greater.

We also evaluate the necessity of policy-based fusion layer. The results show that our policy-based fusion and policy gradient method can improve the accuracy a lot. Also, for Epinions, the improvement is quite significant. One possible reason is that the social effects in Epinions are more heterogeneous than those in WeChat, so to dynamically weigh different social effects are more effective for Epinions. More importantly, since DualEMB can be seen as the extension of DELF \cite{DELF}, where we replace the concatenation of four embedding by our policy unit. Comparing DualEMB in Table \ref{tbl-result2} with DELF in Table \ref{tbl-result1}, we find that the policy unit do help to improve its performance and is superior to straightforward concatenation.

\begin{figure}[h]
	\centering
	\includegraphics[width=0.45\textwidth,angle=0]{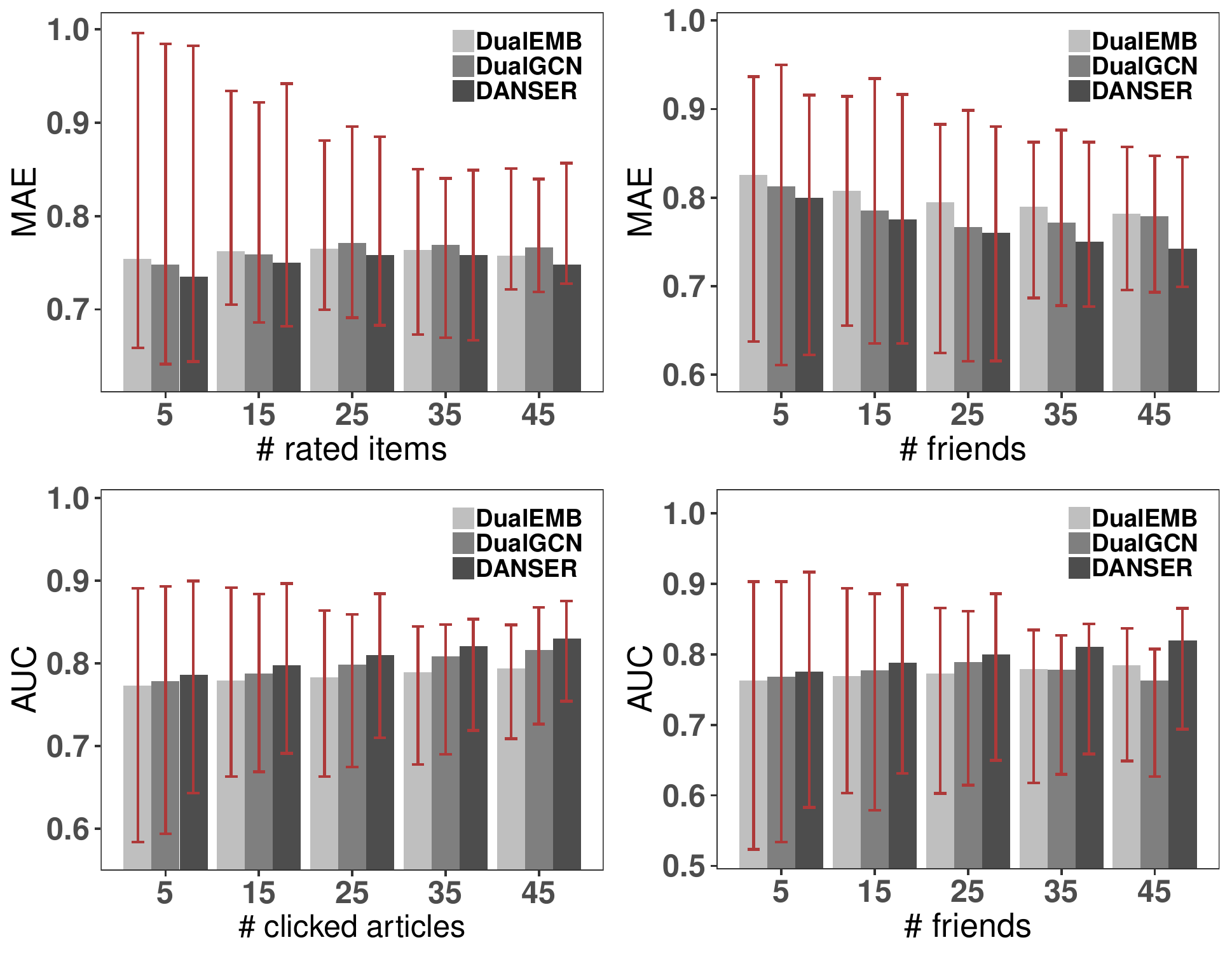}
	\vspace{-5pt}
	\caption{MAE/AUC of DANSER, DualGCN, DualEMB on Epinions/WeChat w.r.t users with different number of clicked items and friends.}
	\label{fig-error}
	\vspace{-10pt}
\end{figure}

In Section 1, we emphasized that social recommendation is proposed to solve the data sparsity and cold-start problem in recommender systems. Here we would like to investigate the performance of DANSER for cold-start users. Fig. \ref{fig-error} shows the results on Epinions and WeChat for users with different number of clicked items and friends. We present the median, the $75-$ and $25-$percentile MAE/AUC. As we can see, with number of clicked items increasing, the gap between upper and lower bounds becomes narrow, which indicates that performance becomes better. Generally, DANSER can achieve better overall performance as well as lower variation of MAE/AUC. For users with different friend numbers, as depicted in the figure, when friend number goes up, the variation trends for three methods are different. For DualEMB, the performance improves a little. For DualGCN, there is an obvious performance decline, while performance of DANSER becomes better. This is possibly because for users with large number of friends, the GCN operation may incorporate more misleading information, and instead, GAT can attentively filter the noise and concentrate more on important neighbor nodes.

%Finally, we evaluate the effectiveness of proposed local-graph aware regularization (LGA Reg) technique, and present the learning curves for train loss, test loss and test MAE in Fig. \ref{fig-train}. The figures show that both our LGA Reg and dropout strategy could help to avoid over-fitting and improve the performance on test set.

\begin{comment}
\begin{figure}[h]
	\centering
	\vspace{-5pt}
	\includegraphics[width=0.45\textwidth,angle=0]{train.pdf}
	\vspace{-5pt}
	\caption{Training curves of DANSER on Epinions for train loss, test loss and test MAE w.r.t different epochs.}
	\label{fig-train}
	\vspace{-10pt}
\end{figure}
\end{comment}

\begin{table}[t]
	\centering
	\caption{Ablation study of components in proposed method.}
	\vspace{-8pt}
	\label{tbl-result2}
	\small
	\begin{threeparttable}
		\begin{tabular}{c|c|c|c|c}
			\toprule[1pt]
			\specialrule{0em}{1pt}{1pt}
			\centering  & \multicolumn{2}{c|}{Epinions} & \multicolumn{2}{c}{WeChat} \\
			\specialrule{0em}{1pt}{1pt}  \cline{2-5}
			\specialrule{0em}{1pt}{1pt}
			~ & MAE & RMSE & P@10 & AUC   \\ 
			\specialrule{0em}{1pt}{1pt} \hline \specialrule{0em}{1pt}{1pt}
			DualEMB  & .7920(1.7\%)\tnote{1} & 1.0363(0.7\%)& .0794(3.6\%)& .7992(2.2\%) \\ 
			DualGCN  & .7840(0.7\%) & 1.0335(0.4\%) & .0814(1.1\%)& .8102(0.8\%)  \\ 
			userGAT & .7858(0.9\%)& 1.0364(0.7\%) & .0813(1.2\%) & .8136(0.4\%)   \\
			itemGAT & .7919(1.7\%) & 1.0335(0.4\%)& .0813 (1.2\%)& .8138(0.3\%)  \\
			\midrule[0.4pt] 
			DANSER-w& .8191(4.9\%)& 1.0659(3.4\%)& .0820(0.4\%) & .8151(0.2\%)  \\
			DANSER-m& .8211(5.2\%)& 1.0681(3.6\%) & .0815(1.0\%)& .8144(0.3\%)  \\
			DANSER-a& .8232(5.4\%)& 1.0710(3.9\%) & .0814 (1.1\%)& .8140(0.3\%)    \\
			DANSER-c& .8091(3.8\%)& 1.0659(3.4\%) & .0809(1.7\%)& .8118(0.6\%) \\
			
			\midrule[0.4pt]
			DANSER & \textbf{0.7787} & \textbf{1.0292} & \textbf{0.0823} & \textbf{0.8165} \\
			\specialrule{0em}{1pt}{1pt}
			\bottomrule[1pt]
		\end{tabular}
		\begin{tablenotes}
	\small
	\item[1] The ratio indicates the impv. comparing DANSER with corresponding variant.
\end{tablenotes}
	\end{threeparttable}
	\vspace{-10pt}
\end{table}

\subsection{Parameter Sensitivity: RQ3}

\begin{figure}[h]
	\centering
	\vspace{-5pt}
	\includegraphics[width=0.45\textwidth,angle=0]{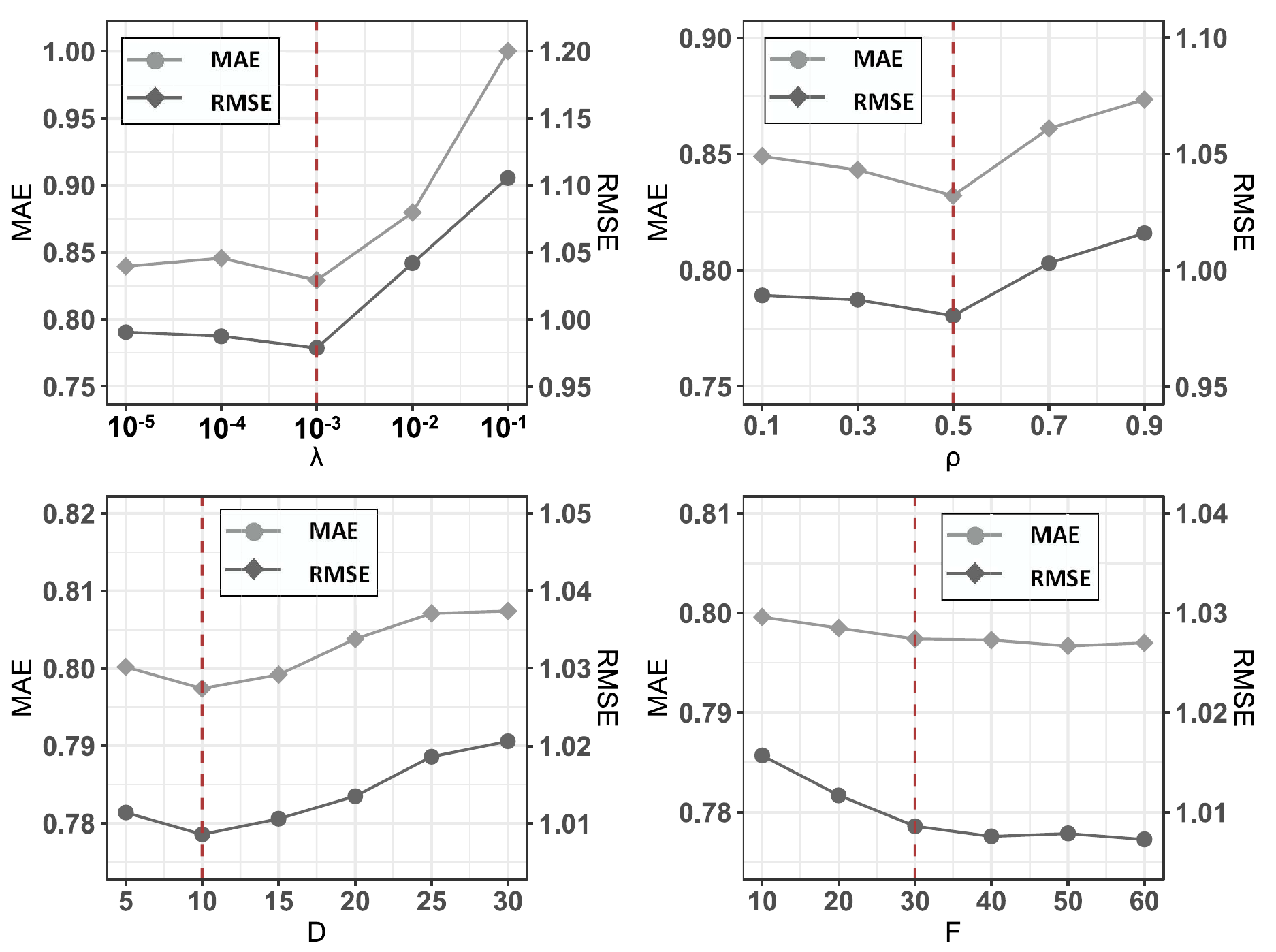}
	\vspace{-10pt}
	\caption{MAE of DANSER on Epinions w.r.t different hyper-parameters. The vertical dotted lines mark the values set in this paper.}
	\label{fig-parameter}
	\vspace{-10pt}
\end{figure}

We study the performance variation for our model w.r.t some hyper-parameters including regularization parameter $\lambda$, dropout rate $\rho$, embedding dimension $D$, sample size $F$. (The other hyper-parameters have little impact on model performance, so we skip discussions on them for space limit.) The results are shown in Fig. \ref{fig-parameter} and enlighten us that. i) The regularization parameter plays an important role, and if it is too large, the model cannot focus on minimizing the recommendation loss. ii) If the dropout ratio is too large, the random pattern would hinder the training. If $\rho$ is too small, it would lose the regularization efficacy. iii) A proper embedding size is needed. If it is too small, the model lacks expressiveness, while if it is too large, the representation vector would become so sparse, which leads to performance decline. iv) With the sample size increasing, the accuracy improves but the computational cost increases as well. Hence an appropriate sample size can keep a good balance between complexity and accuracy.

\subsection{Case Study: RQ4}

\begin{figure}[h]
	\centering
	\includegraphics[width=0.45\textwidth,angle=0]{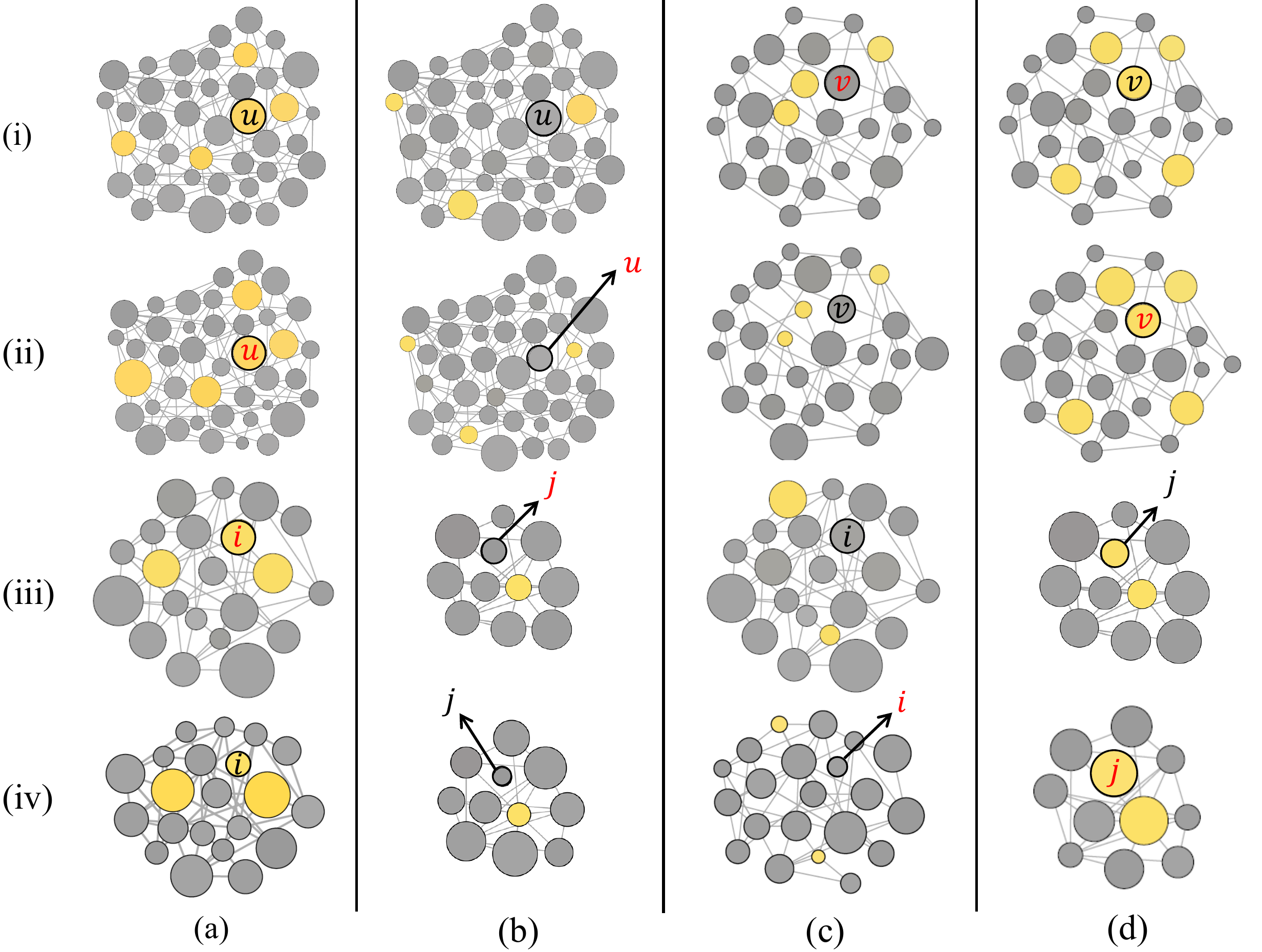}
	\caption{Visualization of the four GATs' weights for a case study. There are four user-item pairs: (a) $(u,i)$, (b) $(u,j)$, (c) $(v,i)$, (d) $(v,j)$. Row (i) and row (ii) are for social homophily and social influence GATs' weights, respectively, while row (iii) and row (vi) are for item-to-item homophily and item-to-item influence GATs' weights, respectively. Gray nodes are negative samples ($r_{ui}=0$) while orange nodes are positive ones ($r_{ui}=1$). In each line, the interacted feature of two factors with red symbol corresponds to the largest active probability given by policy networks.}
	\label{fig-vis} 
	\vspace{-10pt}
\end{figure}

We conduct a case study of four user-item pairs and show the weights given by four GATs by node size in Fig. \ref{fig-vis}. As we can see, the weights of two homophily GATs stay unchanged for the same user or item, while the weights of two influence GATs vary for different user-item pairs. Moreover, influence GATs incline to allocate larger weights to nodes with the same labels, which echoes the argument in Section 3.3 that the influence GATs tend to concentrate more on the similar nodes under a specific context. Also, the policy weights play a part in selecting the most dominant interacted features from four GATs representations. 

More importantly, we can decompose the model interpretability for social effects in recommender systems by a two-level illustration: i) for a specific user-item pair $(u,i)$, the policy weights first indicate the dominant combinations of social effects in user domain and item domain, ii) the corresponding GATs for four social effects further give out the influential degree of each friend given item $i$ as well as the influential degree of each related item given user $u$.

\section{Related Works}

\textbf{Social Recommendation.} In online social networks, friends tend to have a large impact on users' decisions and behaviors \cite{diffuse1, diffuse2}. There are plenty of studies for social effects, such as homophily\cite{Homophily} and influence\cite{Influence}, over graph structures in social network analysis. For recommender systems, \cite{earliest} is one of the pioneer who takes advantage of social networks to alleviate the data sparsity problem for item recommendation. Then many following studies begin to model social effects in recommender systems from various perspectives. \cite{Trust1,Trust2,Trust3} propose trust propagation methods by leveraging rating of friends to deduce rating of targeted user. This kind of approach is quite heuristic and is called \emph{similarity-based} method. Then a more modern approach, \emph{model-based} method, is proposed by other studies. \cite{SocReg,SocialMF} use regularization to constrain the neighbor users' embedding vectors to be similar and \cite{MF1,MF2,TrustMF} adopt matrix factorization to incorporate the adjacent matrix into user embedding. Since the above methods can only capture linear information, some recent studies like \cite{Deep1,NSCR} and \cite{SREPS,NetRep1,DBLP:conf/cikm/ZhangLNLX18} leverage deep neural networks and network embedding approaches, respectively, to learn a more complex representation for graph structures. However, comparing with dual GATs used in this paper, the common limitations of existing studies lie in: i) they assume neighbors' influences to be equally important or statically constrained, ii) they ignore the social effects from related items, iii) modeling of social effects lacks interpretability.

\textbf{Graph Convolution/Attention Network.} GCN and GAT, as two powerful techniques to encode a complex graph into low-dimensional representations, are extensively applied into various problems involved with graph data since their proposals \cite{GCN1, GCN2, GAT}. Using GCN and GAT to solve semi-supervised classification problem in graph could achieve state-the-of-art performance (GAT can be seen as an extension of GCN and provide better performance according to \cite{GAT}), and its good scalability enables it to tackle large-scale dataset. Existing studies leverage GAT to tackle social influence analysis\cite{DeepInf}, graph node classification\cite{Adaptive}, conversation generation\cite{ConGen}, relevence matching\cite{DBLP:conf/cikm/ZhangLNLX18}. For recommendation, some recent studies like \cite{WebScale} adopt GCN to convolve on user-item network (a bipartite graph) to obtain better representations for items, and there are also some works using GCN to capture social information for recommendation \cite{GCNsocial1}\cite{GCNsocial2}. We are the first to use GAT for social recommendation task, and our new architecture, dual GATs, can capture social information in both user and item networks.

\textbf{Dual Mechanism.} There are many dual phenomenons in real-life which inspire several dual structures in model design. \cite{DualModel} proposes a dual-model paradigm to merge the training of two dual tasks. \cite{DGCN} extends GCN to dual structures, combining global and local consistencies in graph. \cite{DELF} proposes DELF conducting dual embedding for users and items in recommender systems. 
{Compared to} \cite{DELF}, our model possesses several key differences: i) for the item-based user embedding (resp. the user-based item embedding), DANSER considers dynamic embedding under specific candidate items (resp. targeted users), while DELF treats embedding statically given a user (resp. item); ii) in dual GCN/GAT layer, DANSER harnesses two dual GATs to capture social information in user social network and item implicit network, while DELF considers no user-to-user and item-to-item mutual influences; iii) in {the} policy-based fusion layer, DANSER dynamically weighs the importances of {the} four GAT representations, while DELF straightforward{ly} concatenate{s} four embedding vectors.

\section{Conclusion}

In this paper, we propose{d} DANSER, {which includes two} dual graph attention networks, to learn deep representations for social effects in recommender systems. Our method is equipped with good expressiveness because: i) it can collaboratively model homophily and influence effects from both friend users and related items; ii) it can investigate heterogeneous importance distributions among different interactions of {the} two two-fold social effects for distinct user-item pairs. Our comparative experiments and ablation studies on a benchmark dataset and a commercial dataset {showed} that DANSER {can} learn effective representations of multifaceted social effects and leverage this information to significantly improve recommendation accuracy.

{As extension of our work, we believe that the construction of item implicit network could leverage extra attribute features or distill more high-level connections through knowledge graphs if the information is available.
More generally, one could extend DANSER to other information retrieval tasks, like question answering, where DANSER could model 'social effects' from related questions and similar answers given the corresponding networks of them.}

\bibliographystyle{ACM-Reference-Format}
\bibliography{sample-sigconf}

\end{document}